\documentstyle[aps,prl,epsfig,preprint]{revtex}
\begin{document}
\title{Dark Area Theorem}
\author{J. H. Eberly$^{a,b}$ and V. V. Kozlov$^{a,c}$}
\address{$^{a}$Rochester Theory Center for Optical Science and Engineering,
         University of Rochester, Rochester, New York 14627\\
         $^{b}$Department of Physics and Astronomy.
         University of Rochester, Rochester, New York 14627\\
         $^{c}$Abteilung f\"{u}r Quantenphysik, Universit\"{a}t Ulm,
          Ulm, Germany, 89081}
\date{\today}
\maketitle
\tightenlines
\begin{abstract}
We report the discovery of a ``dark area theorem," a new
quantum optical relation for propagation of unmatched pulses in
thick three-level $\Lambda$-type media. We define dark area and
derive the dark area theorem for a coherently prepared and
inhomogeneously broadened lambda medium. We also obtain the first
equation for the spatial evolution of the dark state amplitude prior
to pulse-matching.
\end{abstract}

Many advances have been made recently in controlling
quantum systems with a pair of near-resonant optical pulses under
conditions of relatively long-preserved two-photon coherence among
three quantum states. Key parameters being controlled include laser
pulse intensity, frequency, duration, shape, timing and detuning.
Well-known techniques and phenomena such as Rabi splitting
\cite{AutTow}, spectroscopic dark-states \cite{Gozzini-etal}, pulse
matching \cite{Cook-Shore}, and anti-intuitive excitation
\cite{Oreg-etal} have been exploited, with consequences including EIT
(electromagnetically induced transparency \cite{EIT}), LWI (lasing
without inversion \cite{LWI}), state-selective molecular excitation
(STIRAP \cite{STIRAP}), and demonstrations of slow light
\cite{slowlight} and fast light \cite{fastlight}, to mention only a
few prominent examples.

However, in practically every instance a strong field is used to
control a near-resonant medium so that a weak field can penetrate or
amplify or excite transitions in a way normally forbidden. These are
thin absorber applications, in the sense that the strong field's
evolution in the medium is ignored.

The understanding of pulse evolution in two-level media is much more
complete. The McCall-Hahn Area Theorem \cite{MccHah} is available to
provide a unifying picture of the way the medium acts back
self-consistently. Even in situations where coherence is not
complete, the two-level Area Theorem is a guide to the key role of
self-consistent back action, and may suggest inverse applications in
which the medium controls the field \cite{echoes-etal}, i.e. thick
absorber applications.

In the present note we report the discovery of a new nonlinear
propagation law for unmatched pulses in a three-level medium.
This new propagation law, which is appropriately labelled a  Dark
Area Theorem, takes the form:

\begin{equation}\label{DAT}
{\partial\theta_D \over \partial\zeta} = - {\alpha + \alpha' \over 4}
\sin(\theta_D + \phi) + {\alpha - \alpha' \over 4}(\sin\theta_D +
\sin\phi) ,
\end{equation}
where the constant $\phi$ and the ``dark area" $\theta_D$ are defined
below. Here $\alpha$ and $\alpha'$ are the conventional Beer's Law
absorption
coefficients for the two transitions shown in Fig. ~\ref{threelevels}(a).

The fundamental equations, from which the dark area theorem is
derived, are familiar. For the physical pulses we have:

\begin{eqnarray}
&& {\partial \over \partial\zeta}\Omega_{g} = i\mu\langle\, ag^*\rangle\, ,
\label{waveeq1} \\
&& {\partial \over \partial\zeta}\Omega_{g'} = i\mu'\langle\, ag'^* \rangle\, ,
\label{waveeq2}
\end{eqnarray}
where the notation is indicated in Fig.~\ref{threelevels}(a). The
$\Omega$'s are the two Rabi frequencies of the (assumed unchirped)
fields: $\Omega_g \equiv 2d_g{\cal E}_g/\hbar$, etc. The angular
brackets are defined by $\langle
(\dots)\rangle \equiv \int F(\Delta) (\dots) d\Delta$, which is an
average over detunings arising from inhomogeneous broadening, where
$F$ is a normalized distribution, taken symmetric and Gaussian in the
examples calculated below, with $F_0$ setting the time scale for
inhomogeneous relaxation. We take the same $F$ for the two
transitions. The $\mu$ parameters are related to the $\alpha$'s in
the standard way for inhomogeneously broadened media: $\alpha =
\pi\mu F_0$, and $\alpha' = \pi\mu' F_0$, and we do not assume $\mu =
\mu'$. Finally, we have defined local-time coordinates $\tau$ and
$\zeta$ in the frame propagating with velocity $c$ in the medium:
$c\tau \equiv ct - z$ and $\zeta \equiv z$.

In addition to the two field equations there are three equations for
the atomic amplitudes $g$, $g'$ and $a$. We prefer to write these
equations in the dressed ``bright-dark" basis sketched in
Fig.~\ref{threelevels}(b):

\begin{eqnarray}
&& i{\partial \over \partial \tau}D = -{1\over 2}\Omega_{D}^*\, B\, ,
\label{D}\\
&& i{\partial \over \partial \tau}B = -{1\over 2}\Omega_{B}\, a\,
-{1\over 2}\Omega_{D}\, D\,
\label{B}\\
&& i{\partial \over \partial \tau}a = (\Delta-i\gamma)a
-{1\over 2}\Omega_{B}~B,
\label{a}
\end{eqnarray}
where we have defined the bright Rabi frequency as $\Omega_B \equiv
\sqrt{\Omega_{g}^2 + \Omega_{g'}^2}$, and the dark Rabi frequency via
the Fleischhauer-Manka relation \cite{Fleischhauer}: $\Omega_D  =
-i(\dot{\Omega}_g\Omega_{g'} -\dot{\Omega}_{g'}\Omega_g)/\Omega_B^2$.
In these equations and throughout the paper the symbols $B$ and $D$
denote the bright and dark state amplitudes, which are related by a
well-known field-dependent dressing rotation \cite{Kuklinski-etal} in
$g$ - $g'$ space:

\begin{equation}
\left[\begin{array}{c} B \\ D \\ \end{array}\right] =
\left[\begin{array}{cc} \cos\theta/2 & \sin \theta/2  \\ -\sin \theta/2
& \cos \theta/2 \end{array}\right]
\left[\begin{array}{c} g \\ g' \\ \end{array} \right],
\label{eq:dressingRot}
\end{equation}
where $\cos\theta/2 = \Omega_g/\Omega_B$ and $\sin\theta/2 =
\Omega_{g'}/\Omega_B$.

However, we have discovered that the $\pm 2n\pi$ ambiguity of this
definition of $\theta$ has physical meaning, and that the meaning is
related to the spatial evolution of $\theta$. This is a double
surprise - that the ambiguity is physical and that the dressing angle
should be treated as a propagation variable, actually as an ``area"
corresponding to the dark Rabi frequency:

\begin{equation}
\theta (\zeta ,\tau ) \equiv i\int_{-\infty}^\tau
\Omega_D(\zeta, \tau^\prime ) d\tau^\prime +
\theta (\zeta ,-\infty )\, ,
\label{thetaDef}
\end{equation}
where $\tan{1\over 2}\theta (\zeta ,-\infty) \equiv \Omega_{g^\prime}
(\zeta, -\infty ) / \Omega_g (\zeta, -\infty )$.
Now we explain these remarks by examining the previously unexplored
spatial evolution of $\theta$.

The first Dark Area propagation equation is:
\begin{eqnarray}
&& {\partial\theta\over\partial\zeta} = {2i\over \Omega_B}
\Big(\mu_{+}\langle aD^*\rangle -
\mu_{-}\langle aB^*\rangle\Big)\, ,
\label{Area1Eqn}
\end{eqnarray}
where the coefficients are $\theta$-dependent and defined as
$\mu_{+}\equiv {1\over 2}(\mu +\mu^\prime ) + {1\over 2}(\mu^\prime
-\mu)\cos\theta$  and $\mu_{-}\equiv {1\over 2}(\mu
-\mu')\sin\theta$. This equation is derived from the definition of
dark area and equations (\ref{waveeq1}) and (\ref{waveeq2}), and with
it  we
obtain the key to the propagation regime. An expression for $\langle aD^*
\rangle$ and $\langle aB^* \rangle$ can be obtained in two steps,
first by formal integration of the $a$ equation up to a time $T$
following the passage of the pulses, to obtain

\begin{equation}
\label{rho21integral}
a(T;\Delta) = {i\over 2}\int_{-\infty}^T e^{-(\gamma +
i\Delta)(T-\tau)} \Big(\Omega_B ~B \Big) d\tau ,
\end{equation}
and second by substituting this integral into the right side of
(\ref{Area1Eqn}), where the resulting double integrals (over $\tau$
and $\Delta$) can be evaluated \cite{EberlyOE}. This is done in the
regime of rapid inhomogeneous relaxation, in which $F_0$ is a shorter
time than the pulse durations or the medium's homogeneous lifetimes,
and we obtain at time $T$: $i\langle aD^*\rangle \approx -{\pi \over 2}F_0
\Omega_B  D_0 B_0$ and  $i\langle aB^*\rangle \approx -{\pi \over
2}F_0 \Omega_B B_0^2$. The subscript $0$ here denotes $\Delta = 0$.

Now we restrict attention to the case $\mu = \mu'$ for
simplicity in describing the developments, in which case the central
propagation equation (\ref{Area1Eqn}) reduces to:

\begin{equation}
{\partial\theta_D\over\partial\zeta}= {2i\mu\over \Omega_B}
\langle aD^*\rangle  = -\alpha D_0B_0,
\label{Area2Eqn}
\end{equation}
where $\theta_D$ is the total dark area, i.e., $\theta_D =
\theta(\zeta, \tau>T)$. Within the $\tau > T$ regime we apply the
solutions for $B$ and $D$ that are compatible with the asymptotic
preparation of the medium, namely $B = \sin\frac{1}{2}\;(\theta_D +
\phi)$ and $D = \cos\frac{1}{2}(\theta_D + \phi)$, where
$\tan(\phi/2) = g/g'$ taken at $\tau\to -\infty$. This converts
(\ref{Area2Eqn}) directly into the Dark Area Theorem:

\begin{equation}
{\partial\theta_D \over \partial\zeta} =
- {\alpha  \over 2} \sin(\theta_D + \phi),
\label{reducedDAT}
\end{equation}
which is easily seen to be the reduced form of (\ref{DAT}) when
$\alpha = \alpha^\prime$. The more general equation (\ref{DAT}) can
be derived in the same way, simply by retaining $\alpha \ne
\alpha^\prime$ throughout. The Dark Area theorem makes asymptotic
predictions for the pulses, and assigns distinct physical content to
each of the branches of its solution shown in Fig.~\ref{branches}.

As in the two-level case, the three-level area theorem does not
predict pulse shapes, and an infinite variety of shapes can be
asymptotically  stable. In Fig.~\ref{unmatched} we show the evolution
of two initially unmatched and temporally offset pulses, with initial
areas $1.5\pi$ and $4.8\pi$. Propagation over 20 Beer's lengths shows
highly non-trivial shifting and reshaping. In addition one can check
that the final matched pulses conform to the dark area theorem's
asymptotic rule: their amplitude ratio is given by $\tan (\phi /2)$.

A further surprise arising from the use of the dressing angle as a
propagation variable is that the spatial behavior of the bright and
dark amplitudes is already specified by the dressing transformation
itself. In the $\tau > T > F_0$ regime, where the space-time behavior
of $B$ and $D$ is a self-consistent reponse to the pulses, we can
write $\partial B/\partial \zeta = (\partial B/\partial
\theta_D)(\partial\theta_D/\partial \zeta)$, and a similar equation
for $D$ . With the aid of (\ref{Area2Eqn}), we can thereby obtain a
new equation for $D^2$ alone: $\partial D^2 /\partial\zeta = \alpha
D^2(1-D^2)$, with the elementary solution:

\begin{equation}
\ln{D^2(1 - D_{in}^{2}) \over (1 - D^{2})D_{in}^2} =  \alpha (\zeta -
\zeta_{in}),
\label{DSoln}
\end{equation}
where $D_{in}$ here means the value of $D$ at the arbitrary position
of incidence $\zeta_{in}$. The new $D^2$ equation itself already
shows that the alternate solution $D=0$ is not stable, which is
exactly the conclusion reached
in our earlier numerical study of the distinction between $SIT$-type
and $EIT$-type propagation \cite{KozEbe1}.

More important, the analytic solution given here also provides the
first compact predictive expression \cite{EbeRahGro} for spatial
evolution of the dark state toward matched pulses, and confirms what
has been known empirically, that the real pulses finally must become
matched, which is the same as the limit $D^2 \to 1$. The straight
lines of data points given in the inset in Fig.~\ref{DarkPop} show
the agreement between a completely numerical solution obtained from
the original equation set (\ref{waveeq1}) - (\ref{a}) and the simple
expression (\ref{DSoln}) predicted by dark area theory.  The two
different lines of points emphasize that for different times
$\tau/T^*$ in the pulse the spatial evolution of $|D|^2$ will attain
the predicted asymptotic straight-line form more or less quickly.

To summarize, we have presented the explicit form of an area theorem
that is, to the best of our knowledge, the first to be found for
propagation in thick three-level media. The most important ingredient
of the derivation was the discovery that $\theta_D$, twice the
dark-state dressing angle, is the key propagation variable for
physical pulses in coherently prepared three-level media
\cite{Blochangle}. We have demonstrated that, in the domain of rapid
inhomogeneous relaxation, approximate  expressions for the response
of the medium can be found, and that they lead directly to compact
expressions for the  spatial evolution of bright and dark state
amplitudes, as in eqn. (\ref{DSoln}), which were previously unknown.
Numerical solutions show that these simple expressions are highly
accurate.

We expect applications of the dark area theorem to be numerous and
interesting. One can show that the singular points of both
$\tan\theta_D$ and $\cot\theta_D$ determine the number of
zero-crossings of the physical pulse envelopes $\Omega_g$ and
$\Omega_{g'}$, such as the two that appear spontaneously in
Fig.~\ref{unmatched} around $\alpha L = $ 5 and 10. Such
zero-crossings appear closely related to phase jumps previously
observed in Raman solitons \cite{Druhl-etal}. The
inclusion of inhomogeneous broadening makes the dark area theorem
capable of addressing three-level echo effects \cite{Hartmann}.
Application of the singularity rules mentioned above will lead to non-trivial
analogs of the McCall-Hahn rule for split-up of $2n\pi$ pulses. In
conclusion, we mention again that the full expression of the theorem
in eqn. (\ref{DAT}) allows $\mu \ne \mu'$. A detailed  examination of
(\ref{DAT}) must be given elsewhere \cite{KozEbe3},  along with the
spatial development of the bright field $\Omega_B$ and  other
elements missing here for lack of space.

Acknowledgement: Research partially supported by NSF grants PHY94-15583 and
PHY00-72359, and the programme QUIBITS of the European Commission.

\newpage
\section*{Figure Captions}
\begin{figure}[htb]
\begin{center}
\epsfig{file=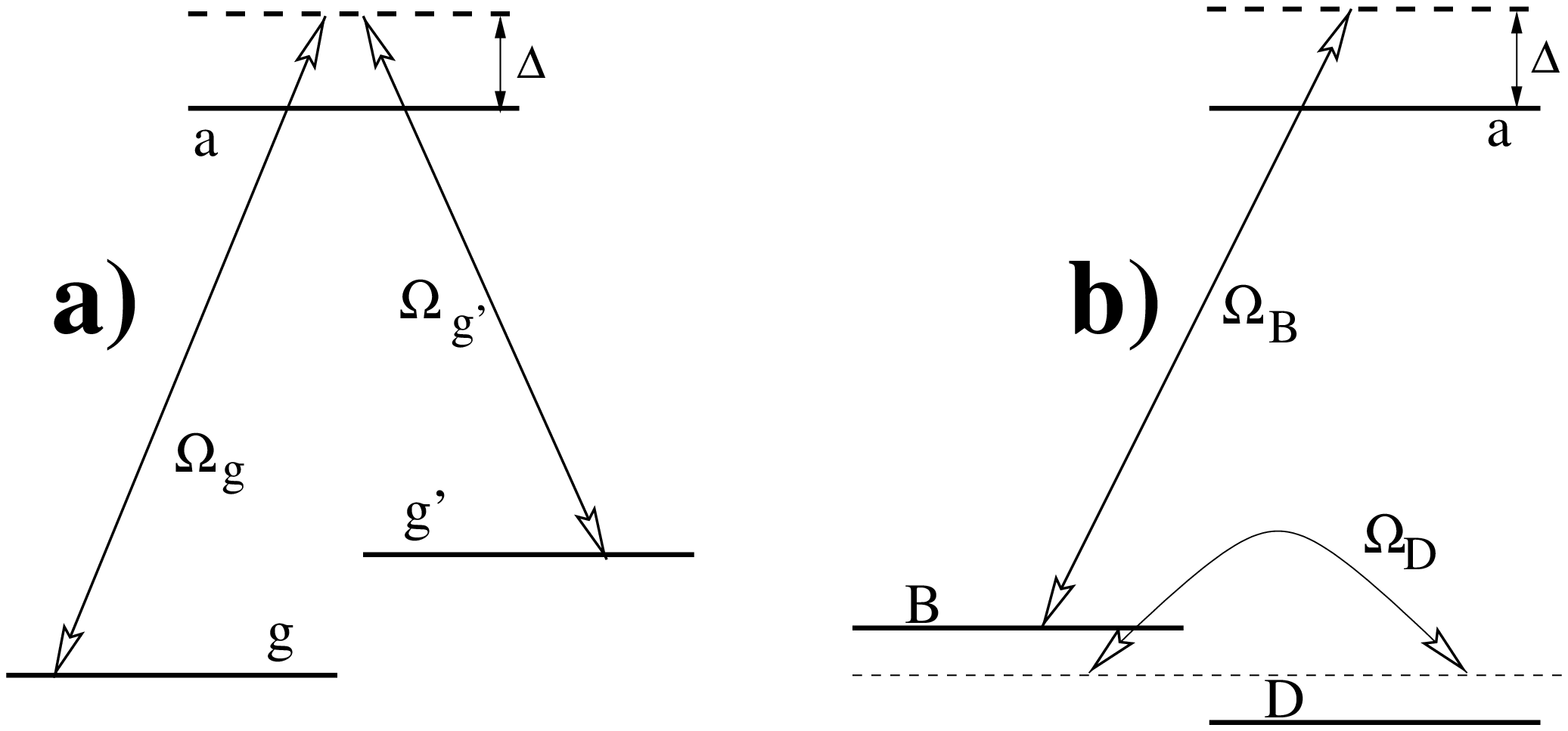, width=6cm, angle=270}

\bigskip
\caption{\label{threelevels}
J.H. Eberly and V.V. Kozlov, ``Dark Area Theorem''}
\end{center}
\end{figure}

Fig.~\ref{threelevels}\\
a) Excitation of the bare-state atom by two pulsed fields,
denoted by their Rabi frequencies $\Omega_g$ and $\Omega_{g'}$, and
b) excitation of the counterpart dressed-state atom by bright and
dark fields, as defined in the text below eqn. (\ref{a}).

\begin{figure}[htb]
\begin{center}
\epsfig{file=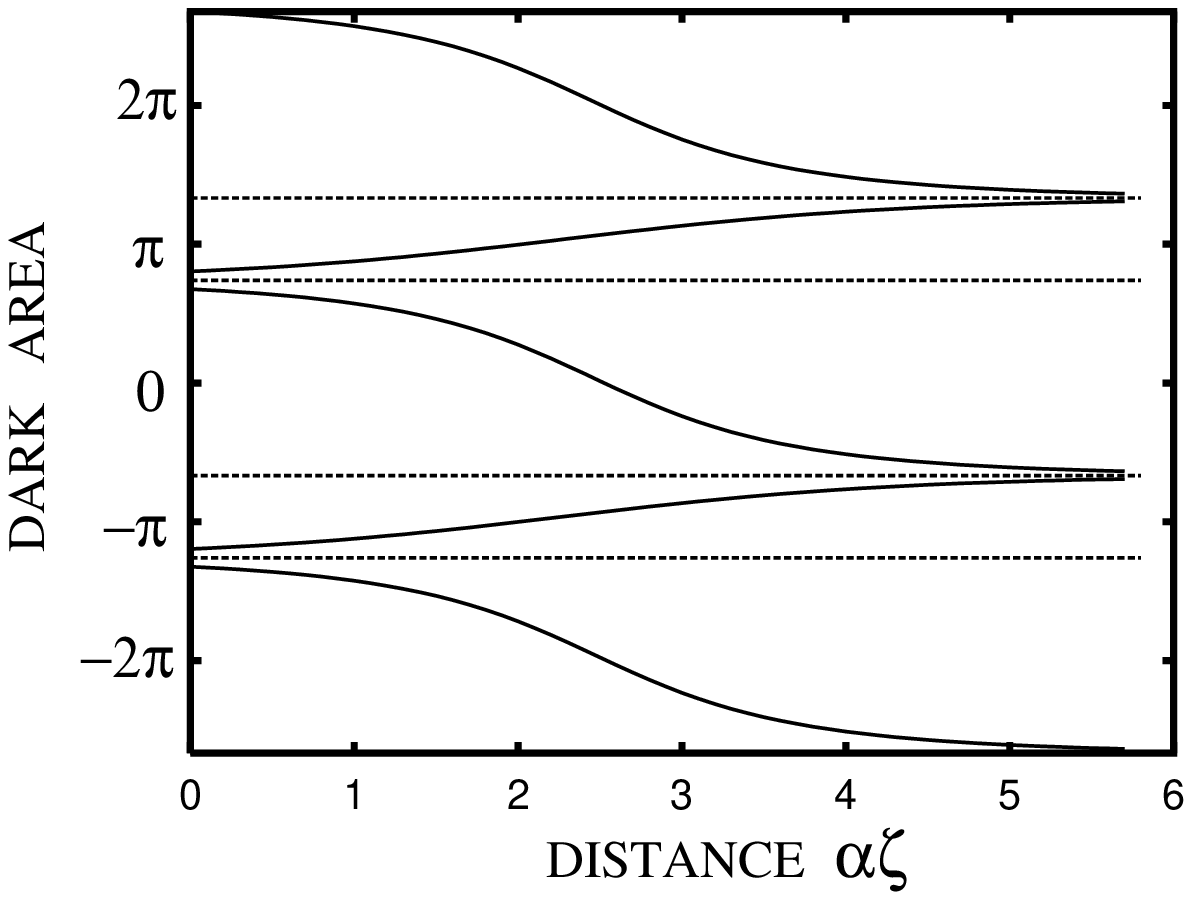, width=10.0cm, angle=270}

\bigskip
\caption{\label{branches}
J.H. Eberly and V.V. Kozlov, ``Dark Area Theorem''}
\end{center}
\end{figure}

\bigskip
Fig.~\ref{branches}\\
Evolution of the dark area $\theta_D$ with distance, as
predicted by eqn. (\ref{DAT}). Asymptotes are determined by $\theta_D
+ \phi = 2n\pi$ and $\tan(\theta_D/2)\tan(\phi/2) = \alpha/\alpha'$.
The values used here are $\phi = 2\pi/3$ and $\alpha' = 4\alpha$.

\begin{figure}[htb]
\begin{center}
\epsfig{file=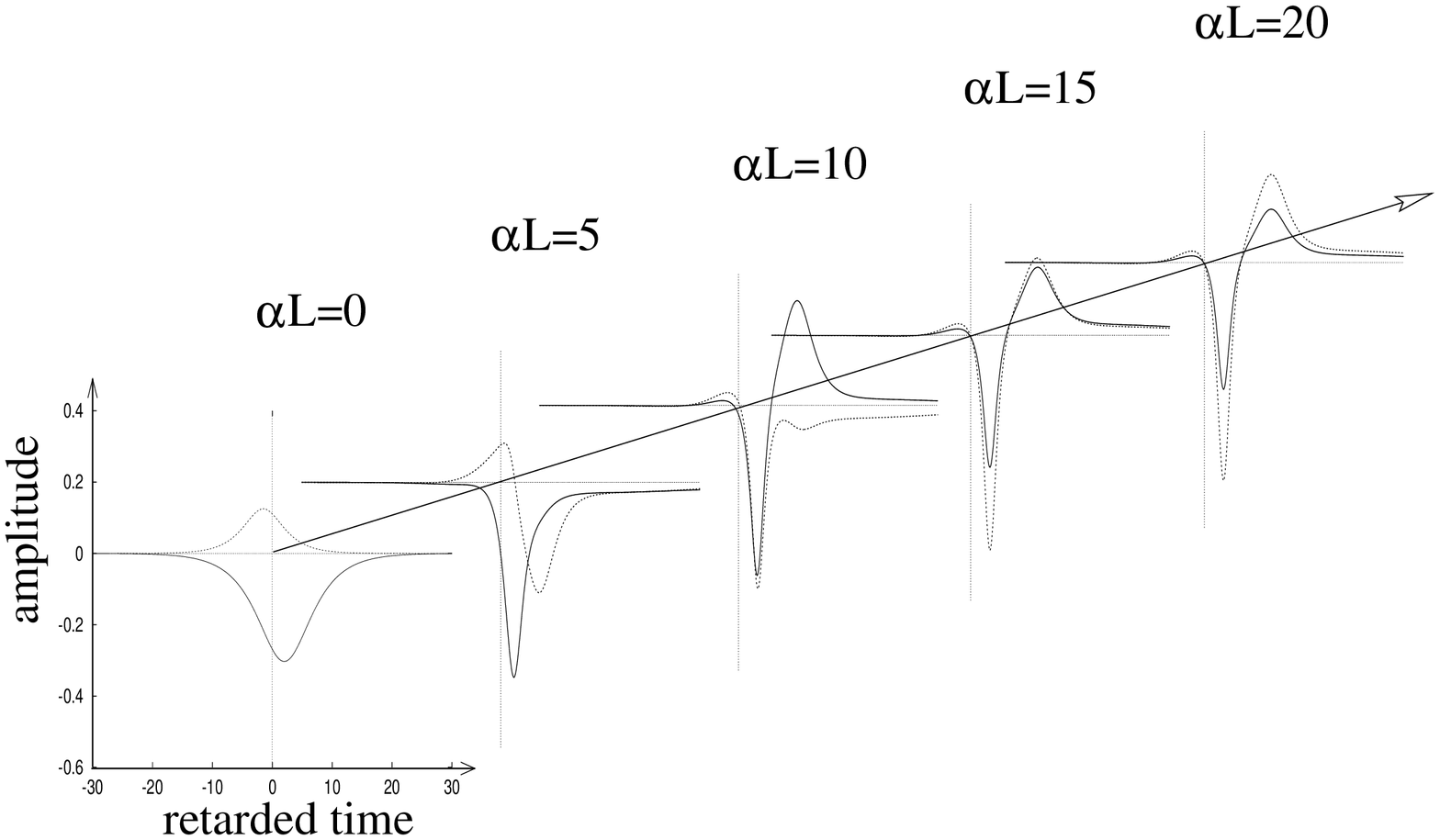, width=8.0cm, angle=270}

\bigskip
\caption{\label{unmatched}
J.H. Eberly and V.V. Kozlov, ``Dark Area Theorem''}
\end{center}
\end{figure}

\bigskip
Fig.~\ref{unmatched}\\
Spatial snapshots of the evolution of two initially
unmatched pulses with areas $1.5\pi$ and $4.8\pi$, through 20 Beer's
lengths of a $\Lambda$ medium with $\phi = 2\pi/3$, obtained by exact
numerical solution of eqns. (\ref{waveeq1})-(\ref{a}), as a test of eqn.
(\ref{reducedDAT}). The inhomogeneous detuning distribution 
$F(\Delta)$ is taken Gaussian, and local time is shown in units of 
$T^* = \sqrt{2\pi}F_0$. At $\alpha\zeta = 20$ the pulses are almost 
matched, with $\Omega_{g'}/\Omega_{g} \approx 1.6$, in good agreement 
with the final amplitude ratio predicted by the branch value of 
$\theta_D$, i.e., $\tan(2\pi - \phi) = \tan(4\pi/3) = \sqrt{3}$.

\begin{figure}[htb]
\begin{center}
\epsfig{file=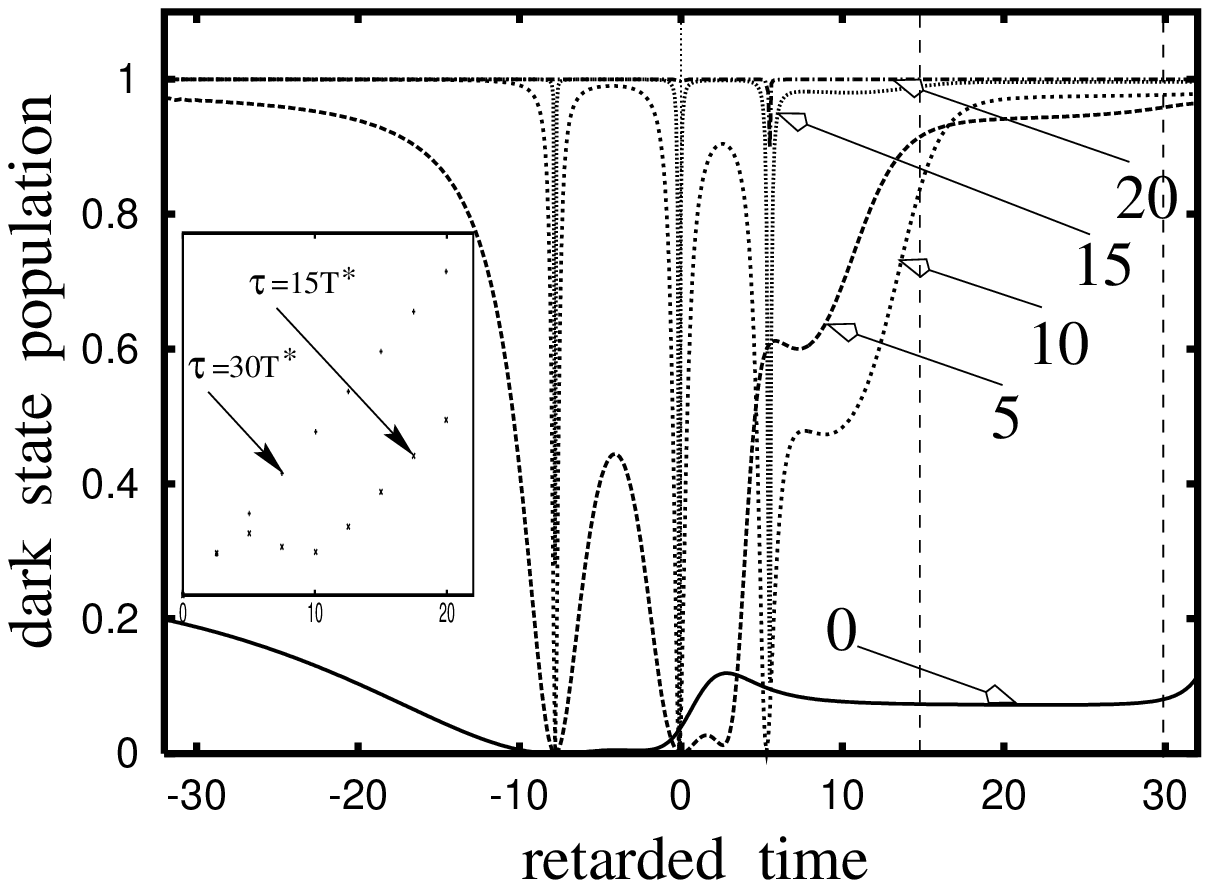, width=10.0cm, angle=270}

\bigskip
\caption{\label{DarkPop}
J.H. Eberly and V.V. Kozlov, ``Dark Area Theorem''}
\end{center}
\end{figure}

\bigskip
Fig.~\ref{DarkPop}\\
Dark state population as a function of time is shown for
five different depths of propagation of the same pulses shown in
Fig.~\ref{unmatched}. Evolution to $|D|^2 = 1$ is evident. The
specific prediction of eqn. (\ref{DSoln}) is also checked  by
computing $|D(\zeta,\tau)|^2$ for these pulses over 8 equal intervals at
$\alpha\zeta = $ 0, 2.5, 5.0, \dots , 20.0 at both times $\tau/T^* =
$ 15 and 30, indicated by vertical dashed lines in the right-hand
half. The two sets of straight-line data with approximately unit
slope shown in the inset provide excellent confirmation of the dark
area theory.

\begin{thebibliography}{99}
\bibitem{Blochangle} Recall that in dressed-state treatments of
two-level interactions the dressing angle is half the Bloch-sphere
angle $\theta$, which is of course also the two-level area.
\end{thebibliography}
\end{document}